\documentclass[a4paper,12pt]{article}
\usepackage{jcappub}
\usepackage[applemac]{inputenc}
\usepackage{graphicx}
\usepackage{amsmath}
\usepackage{amsfonts}
\usepackage{amssymb}
\usepackage{makeidx}
\usepackage{mathtools}
\usepackage{wrapfig}
\usepackage{soul}
\usepackage{empheq}
\usepackage{geometry}
\usepackage{stackengine}
\newcommand{\be}{\begin{equation}}
\newcommand{\ee}{\end{equation}}

\newcommand{\bea}{\begin{eqnarray}}
\newcommand{\eea}{\end{eqnarray}}

\newcommand{\HH}{\mathcal{H} }

\newcommand{\R}{\mathcal{R} }

\newcommand{\ndv}{{v_{||}}}

\newcommand{\bn}{{ \bf n }}
\newcommand{\bx}{{ \bf x }}
\newcommand{\bk}{{ \bf k }}
\newcommand{\bq}{{ \bf q }}
\newcommand{\bv}{{ \bf v }}

\usepackage{hyperref}

 \usepackage{color}

\newcommand{\LCDM}{$\Lambda$CDM}

\newcommand{\thetadm}{\theta_{\rm dm}}
\newcommand{\deltade}{\delta_{\rm de}}
\newcommand{\thetade}{\theta_{\rm de}}

\newcommand{\dd}{{\rm d}}

\newcommand{\rhode}{\rho_{\rm de}}
\newcommand{\rhodm}{\rho_{\rm dm}}

\title{A smoking gun from the power spectrum dipole for elastic interactions in the dark sector}
\author[a]{Jose Beltr\'an Jim\'enez,}
\author[b]{Enea Di Dio}
\author[a]{and David Figueruelo}

\affiliation[a]{Departamento~de~F{\'i}sica~Fundamental~and~IUFFyM,~Universidad~de~Salamanca,~E-37008~Salamanca,~Spain.}
\affiliation[b]{CERN, Theory Department, CH-1211 Geneva 23, Switzerland.}

\emailAdd{jose.beltran@usal.es, enea.didio@cern.ch, davidfiguer@usal.es}

\abstract{Models with pure momentum exchange in the dark sector have been shown to provide a promising scenario to tackle the tension in the clustering inferred from high- and low-redshift probes. A distinctive feature of these models is that only the Euler equation for the dark matter component is modified and the correction is such that the net effect can be associated to an additional friction determined by the interaction rate. In this work, we show that the strength of the interaction parameter needed to resolve the $\sigma_8$ tension could be detected from the dipole of the matter power spectrum that is expected to be measured in upcoming surveys.}

\begin{document}

  \begin{minipage}{.45\linewidth}
    \begin{flushleft}
    \end{flushleft}
  \end{minipage}
\begin{minipage}{.45\linewidth}
\begin{flushright}
 {CERN-TH-2022-192}
 \end{flushright}
 \end{minipage}

\maketitle

\section{Introduction}

The standard model of cosmology, dubbed \LCDM, has shown an excellent agreement with most cosmological observations to date. However, as the amount and precision of the available data  have increased over the last decades, some tensions have surfaced. If these tensions were statistical flukes, we would expect them to disappear, within the cosmic variance limit, as we collect more data. However, not only they do not disappear but, in some cases like the Hubble tension, they even become more pressing~\cite{Planck:2018vyg,Riess:2021jrx}. Among the existing tensions (with an admittedly mild statistical significance), there is an apparent discrepancy between the amplitude of the fluctuations today as extrapolated from  the Cosmic Microwave Background (CMB) measurements~\cite{Planck:2018vyg,ACT:2020gnv} and those derived from low redshift surveys (shear weak lensing~\cite{HSC:2018mrq,DES:2021wwk,Busch:2022pcx}, CMB lensing~\cite{Krolewski:2021yqy,White:2021yvw,DES:2022urg} and galaxy clustering~\cite{DAmico:2019fhj,Ivanov:2019pdj,Kobayashi:2021oud}). This is the so-called $\sigma_8$ tension and might signal towards a possible weaker clustering than that predicted by \LCDM{} and, thus, to the existence of new physics. 

Of course, the $\sigma_8$ tension could be caused by some unknown systematics, but the variety and consistency of the probes where the tension is observed calls for a scrutiny of possible physical mechanisms able to explain the discrepancy and this has paved the way for numerous attempts in this direction. A popular approach consists in modifying the dark sector, either by adding some features to the dark matter component, or by modifying the dark energy sector that indirectly affects the matter clustering or by including some type of interaction within the dark sector. In this work we will be concerned with the latter and, in particular, with a particular class of models with an interaction designed so that the background cosmology remains unaltered and only the Euler equations are affected, i.e., up to linear order in perturbation theory only momentum can be transferred between the dark components. Since there is no energy exchange, these interactions are elastic up to that order.\footnote{Non-linear terms can introduce energy exchange, thus spoiling the elasticity of the interaction. Although this might be relevant for numerical simulations, we only consider the linear regime so for our purposes the interaction is perfectly elastic.} These models have been shown to exhibit some interesting virtues, including a promising scenario to alleviate the $\sigma_8$ tension~\cite{Simpson:2010vh,Pourtsidou:2016ico,Asghari:2019qld,Linton:2021cgd,Chamings:2019kcl,Figueruelo:2021elm,BeltranJimenez:2021wbq,Kumar:2017bpv}.

A very intriguing feature of these models is that the inclusion of measurements of $S_8$ as a Gaussian prior in the fit to data, not only improves the fit with respect to \LCDM, but it leads to a possible {\it detection} of the interaction. This feature was observed in~\cite{Pourtsidou:2016ico} for a model of coupled quintessence with pure momentum exchange. In~\cite{Asghari:2019qld} a model where dark matter and dark energy are coupled via a (covariant) interaction driven by the relative velocity of the corresponding components was analysed and a similar result was hinted. This model was later explored in more detail in~\cite{Figueruelo:2021elm} and~\cite{BeltranJimenez:2021wbq} where it was confirmed the crucial role played by the inclusion of $S_8$ measurements and that, in this case, a {\it detection} of the interaction could be inferred. The results for this model have also been independently confirmed in~\cite{Poulin:2022sgp}. Similar conclusions are also achieved when the baryonic sector interacts with dark energy~\cite{Jimenez:2020ysu}. Finally, a model of pure momentum exchange between dark matter and dark energy has been constructed in~\cite{BeltranJimenez:2020qdu} by using the  Schutz-Sorkin Lagrangian formalism and similar results were also obtained for this scenario in~\cite{BeltranJimenez:2021wbq}. All these previous results make these scenarios worth being further explored and the goal of this work is to unveil potential distinctive signatures of these models that could eventually be detected. In particular, we will study if future measurements of the galaxy power spectrum dipole could confirm a non-vanishing value of the interaction parameter. To this aim, our approach will be as follows: Since existing data (with the caveat of using $S_8$ measurements in mind) signal to the presence of the interaction, excluding the non-interacting case at several sigmas, we will assume a fiducial model with the interaction. Then, we will derive the effect of this interaction in the dipole of the matter power spectrum and show the potential of future SKA-like data~\cite{Braun:2015zta} to detect such an effect.

The work is organised as follows: In Sec.~\ref{sec:model} we introduce the class of models that we will analyse. We will then review the relativistic effects for the galaxy clustering~\ref{sec:rel_effects} and  their relation with the Euler equation~\ref{sec:deviation_Euler}. Then in Sec.~\ref{sec:dipole} we study the particular signature of this class of models in the dipole of the galaxy power spectrum. In Sec.~\ref{sec:results} we present the numerical results taylored for a SKA-like survey and in Sec.~\ref{sec:conclusions} we draw our conclusions.

\section{A proxy for elastic dark interactions}
\label{sec:model}

We will start by introducing the model under consideration in this work. Let us stress that, although we will work with a specific model, our results are expected to hold for the general class of pure momentum exchange models able to alleviate the $\sigma_8$ tension, so we consider our specific scenario as a proxy for this class of models. For reasons that will become clear soon, we will refer to it as covariantised dark Thomson-like scattering. This model was first considered in Ref.~\cite{Asghari:2019qld} and subsequently explored in~\cite{Figueruelo:2021elm,BeltranJimenez:2021wbq,Poulin:2022sgp,Cardona:2022mdq}. The philosophy behind this model is to design an interaction in the dark sector that only affects the linear perturbations, but leaves the background cosmology intact. This is straightforwardly achieved by introducing an interaction at the level of the energy-momentum conservation equations that is governed by the relative velocities of the involved fluids. By invoking the cosmological principle, such an interaction drops from the background evolution. More explicitly, the interaction is introduced as
\begin{equation}
    \nabla_\mu T^{\mu \nu}_{\rm dm}=Q^\nu\;,\qquad 
    \nabla_\mu T^{\mu \nu}_{\rm de}=-Q^\nu\;, \quad 
    \textnormal{with} \quad Q^{\nu}=\bar{\alpha}\left(u^\nu_{\rm de}-u^\nu_{\rm dm}\right)\;,
    \label{eq:non_conserved_Tmunu}
\end{equation}
where $\bar{\alpha}$ measures the strength of the interaction. In principle, it could depend on both space and time, but we will assume it to be constant. Furthermore, since it has dimension 5, we will instead use the dimensionless quantity 
\begin{equation}
\alpha=\frac{8  \pi G}{3 H_0^3} \bar{\alpha}\;,
\label{eq:normalisation_alpha}
\end{equation}
where the powers of $H_0$ and $G$ have been chosen for convenience. This interaction clearly fulfills our requirements because all the cosmological fluids share the same large scale rest frame so $Q^\mu$ identically vanishes for the background evolution. For the perturbed sector, peculiar velocities appear and the effects from the interaction begin to be relevant. Thus, let us consider a perturbed  Friedmann-Lema\^itre-Robertson-Walker (FLRW) metric described by the line element
\be
\dd s^2 =  a^2(t) \left[ - \left( 1+ 2 \Phi \right) \dd t^2 + \left( 1- 2 \Phi \right) \dd\bx^2 \right]\;,
\ee
where we have fixed the Newtonian gauge and used the absence of anisotropic stresses so the two gravitational potentials are the same.\footnote{Since we work to linear order in perturbation theory and we neglect neutrinos, we can set the two Bardeen potentials to be equal from the beginning.} The perturbed conservation equations read
\begin{eqnarray}
\label{eq:deltaDM_alpha}
\dot{\delta}_{\rm dm}&=&-\thetadm + 3 \dot{\Phi} \;, \\
\dot{\theta}_{\rm dm}&=&-\mathcal{H} \thetadm + k^2 \Phi + \Gamma(\thetade-\thetadm)\;,\label{eq:thetaDM_alpha} \\
\dot{\delta}_{\rm de}&=&-3 \mathcal{H}\left(c_{\rm s}^2-w\right)\deltade +3(1+w)\dot{\Phi}  -\thetade(1+w)\left(1+9 \mathcal{H}^2\frac{c_{\rm s}^2-w}{k^2}\right) \;, \label{eq:deltaDE_alpha}\\
\dot{\theta}_{\rm de}&=&\left(-1+3c_{\rm s}^2\right) \mathcal{H}\thetade+k^2\Phi +\frac{k^2 c_{\rm s}^2}{1+w}\deltade-\Gamma R(\thetade-\thetadm)\;,\label{eq:thetaDE_alpha}
\end{eqnarray}
where we denote with a dot the derivative with respect to the conformal time $t$, $\theta$ stands for the divergence of the peculiar velocity, $w$ and $c_{\rm s}^2$ denote the equation of state and the sound speeds of dark energy respectively, 
$\Gamma$ is the interaction rate between dark energy and dark matter and $R$ is the dark matter-to-dark energy ratio, both defined as 
\begin{eqnarray}
\Gamma&\equiv& \bar{\alpha}\frac{a}{\rhodm} \;, \\ \label{eq:Scoupling_alpha}
R&\equiv&\frac{\rhodm}{(1+w)\rhode} \;. \label{eq:Rcoupling_alpha}
\end{eqnarray}
The above equations, together with the usual Poisson equation that is also unaffected by the interaction\footnote{The evolution of the gravitational potential is of course indirectly affected.}, are the sets of equations governing the evolution of the perturbations and, in particular, the clustering of dark matter. We see that, at the level of linear perturbations, the interaction modifies the Euler equations in the same form as a Thomson scattering (in fact, this term in the perturbation equations was obtained by introducing a scattering between dark energy and dark matter in~\cite{Simpson:2010vh}), hence the name covariantised Thomson-like interaction (since at higher order in perturbation theory differences appear). This interaction is the responsible for erasing structures since it transfers pressure support from dark energy to dark matter and, therefore, structure formation is less efficient. Furthermore, since $\Gamma\propto a^4$, this effect tends to appear at low redshift when  $\Gamma$ is sufficiently large. This is the main mechanism at work that makes this model able to alleviate the $\sigma_8$ tension as shown in the literature~\cite{Asghari:2019qld,Figueruelo:2021elm,Poulin:2022sgp}. We refer to those references for a more detailed explanation and analysis of the model. Here we only want to stress that a value of $\alpha\simeq 1$ is strongly favoured by data when including measurements of $S_8$ as a Gaussian prior. Although there might be some caveats of including  these measurements, it is intriguing that not only the fit with respect to \LCDM{} is improved, but the non-interacting case is excluded at more than 3 sigmas. In the following we want to explore to what extent SKA-like surveys will be sensitive to the presence of this interaction through the dipole of the galaxy power spectrum. Let us then proceed to see how the dipole can shed light on this interaction.

\section{Relativistic description of galaxy clustering}
\label{sec:rel_effects}

Current and upcoming galaxy surveys~\cite{Aghamousa:2016zmz,Laureijs:2011gra,Spergel:2013tha,Dore:2014cca,Schlegel:2019eqc} map the 3-dimensional positions (line-of-sight direction $\bn$ and redshift $z$) of tens of millions of galaxies covering a large fraction of the sky. From these catalogs we can measure the perturbation of the galaxy number density, i.e.~the galaxy number counts
\be
\Delta \left( \bn , z \right) = \frac{N \left( \bn, z \right)- \langle N \rangle \left(z \right)}{\langle N \rangle \left(z \right)}
\ee
where $\langle .. \rangle$ denotes the angular average at fixed observed redshift $z$. The galaxy number counts has been derived in a full relativistic framework in Refs.~\cite{Yoo:2009au,Yoo:2010ni,Bonvin:2011bg,Challinor:2011bk} to first order in perturbation theory and then extended to second order in Refs.~\cite{Yoo:2014sfa,Bertacca:2014dra,DiDio:2014lka}.
In this work we are interested in the linear relativistic terms proportional to the peculiar velocity of galaxies. Indeed, as it has been pointed out in Refs.~\cite{McDonald:2009ud,Bonvin:2013ogt,Croft:2013taa,Bonvin:2014owa,Bonvin:2015kuc,Lepori:2017twd,Breton:2018wzk,DiDio:2018zmk,Beutler:2020evf,DiDio:2020jvo,Tutusaus:2022cab,Sobral-Blanco:2022oel} these terms will lead to  a non-vanishing imaginary part of the cross-power spectrum or, equivalently, they source the dipole of the power spectrum or 2-point correlation functions for two different galaxy populations.
In this approximation (and neglecting integrated terms) the galaxy number counts is described by\footnote{The power counting is in terms of spatial derivatives, which in Fourier space become factors $\HH/k$. For sake of simplicity we use the same notation in real and Fourier space.}
\bea \label{eq:numbercounts}
\Delta  &=& b \delta + \HH^{-1} \partial_r \ndv 
 \nonumber\\
&&+ \HH^{-1} \partial_r \Phi - \left( 1 - 5s + \frac{5s -2}{r \HH} - \frac{\dot \HH}{\HH^2} + f_{\rm evo} \right) \ndv - \HH^{-1} \dot \ndv  + \mathcal{O} \left( \HH^2/k^2 \right)  \, , \nonumber\\
\eea
where $\ndv = \bn \cdot \bv$ and $\bn$ is the unit vector pointing from the source to the observer. In order to relate dark matter fluctuations with the observed galaxies we need to introduce three bias parameters: the galaxy bias $b$, the magnification bias $s$ and the evolution bias $f_{\rm evo}$. The latter are defined as
\bea
s&=& \left. -\frac{2}{5} \frac{\partial \ln n}{\partial \ln L} \right|_{L = \bar L} \, ,
\\
f_{\rm evo} &=& \frac{\partial \ln n}{\partial \ln a} \, ,
\eea 
where $n$ denotes the comoving galaxy density.
These two bias parameters capture the effects of incomplete galaxy sample (in magnitude) and non-trivial distribution of the galaxies in redshift due to galaxy formation, respectively. 
In particular a volume limited survey is described by $s=0$, while $f_{\rm evo} =0$ is obtained when the number of sources is conserved in a comoving volume.
The first line of eq.~\eqref{eq:numbercounts} denotes the standard terms in Newtonian approximation, while the second line encodes the leading relativistic corrections.

In $\Lambda$CDM, all the matter particles move according to the Euler equation 
\be \label{eq:Euler}
\dot \bv + \HH  \bv + \nabla \Phi = 0 \, ,
\ee
or, by contracting with the direction $\bn$, with a radial peculiar velocity satisfying
\be
\label{eq:Eulerradial}
\dot{v}_\parallel + \HH   \ndv - \partial_r \Phi = 0 \, .
\ee
As a consequence, the galaxy number counts to linear order is not sensitive neither to the acceleration $\dot{v}_\parallel$ nor to the gravitational redshift $\partial_r \Phi$ and is given by
\bea \label{eq:numbercounts_LCDM}
\Delta^{\Lambda{\rm CDM}}  &=& b \delta + \HH^{-1} \partial_r \ndv 
\nonumber \\
&&- \left( 1 - 5s + \frac{5s -2}{r \HH} - \frac{\dot \HH}{\HH^2} + f_{\rm evo} \right) \ndv  + \mathcal{O} \left( \HH^2/k^2 \right)  \, .
\eea
In this scenario, we can consider a single galaxy as a test particle whose motion is described by the geodesic equations. It is well-known that, in a matter dominated era and in $\Lambda$CDM, the Euler equation describes the geodesic motion. 
We leave its derivation in Appendix~\ref{app:geodesic}. This will become useful beyond $\Lambda$CDM.
However, as we will see in the next session, the equivalence between the geodesic equation and the Euler equation is broken once we have interacting dark matter, even in the matter dominated era ant this will be the main effect that will make it possible to use the matter power spectrum dipole a smoking gun for the elastic interactions.

\section{Deviation from Euler equation}
\label{sec:deviation_Euler}

Let us consider the modified dark matter Euler equation presented in the previous section
\begin{equation}
    \dot{\theta}_{\rm dm}=-\mathcal{H} \theta_{\rm dm} + k^2 \Phi + \Gamma(\theta_{\rm de}-\theta_{\rm dm}) \;.
\end{equation}
For small enough scales, where precisely the interaction is efficient, and for realistic values of the coupling $\alpha\sim\mathcal{O}(1)$, the dark energy velocity is negligible compared to the dark matter one. Then, we can approximate the Euler equation for dark matter as
\begin{equation}
 \dot{\theta}_{\rm dm}\simeq-\mathcal{H} \theta_{\rm dm} + k^2 \Phi - \Gamma \theta_{\rm dm}\;.
\end{equation}
Consequently, the interaction between dark matter and dark energy discussed above leads to the following modified Euler equation for the dark matter field:
\be
\dot \bv + \HH \left( 1+ \Theta \right) \bv + \nabla \Phi = 0\;,
\ee
or, by contracting with the direction $\bn$, we obtain
\be \label{eq:Eulerradial_theta}
\dot \ndv + \HH \left( 1+ \Theta \right)  \ndv - \partial_r \Phi = 0 \,,
\ee
where the deviation from the standard scenario is encoded in the variable $\Theta$ defined as 
\be
\Theta\equiv\Gamma/\mathcal{H}=\alpha\frac{H_{0} a}{\Omega_{\rm dm}(a)\mathcal{H}}\, .
\ee
This modified Euler equation describes the dark matter motion in the interacting scenario considered in this work and for the relevant scales. Since we observe only galaxies, and the velocity appearing in the galaxy number counts~\eqref{eq:numbercounts} is the galaxy peculiar velocity some assumptions need to be made to study the effects of the interaction. In order to determine the galaxy velocity we will consider the following two possible extreme cases: 
\begin{itemize}
    \item Dark matter tracers: Galaxies comove with dark matter (see e.g.~\cite{Bonvin:2018ckp,Bonvin:2020cxp,Castello:2022uuu}), i.e., galaxies are faithful tracers of the dark matter velocity field.
    \item Gravitational potential tracers: Galaxies move according to the linear geodesic equation so they move according to the gravitational potential (see e.g.~\cite{Koyama:2009gd}).
\end{itemize}
We will analyse these two scenarios for completeness. However, we believe that galaxies will be good tracers of the dark matter velocity field, so we expect the first scenario to be more realistic. The reasoning is that the bulk of structure formation takes place when the interaction is not active so galaxies are locked at the bottom of the dark matter haloes gravitational potentials. Since the realistic values of $\alpha$ gives rise to a small effective friction for dark matter, we do not expect this to be sufficient to drag galaxies away from the potential wells. This issue is being currently investigated through numerical simulations and the results will be presented elsewhere.

\subsection{Dark matter tracers}
\label{sec:same_vel}
\begin{figure}[ht!]
\centering
\includegraphics[scale=0.20]{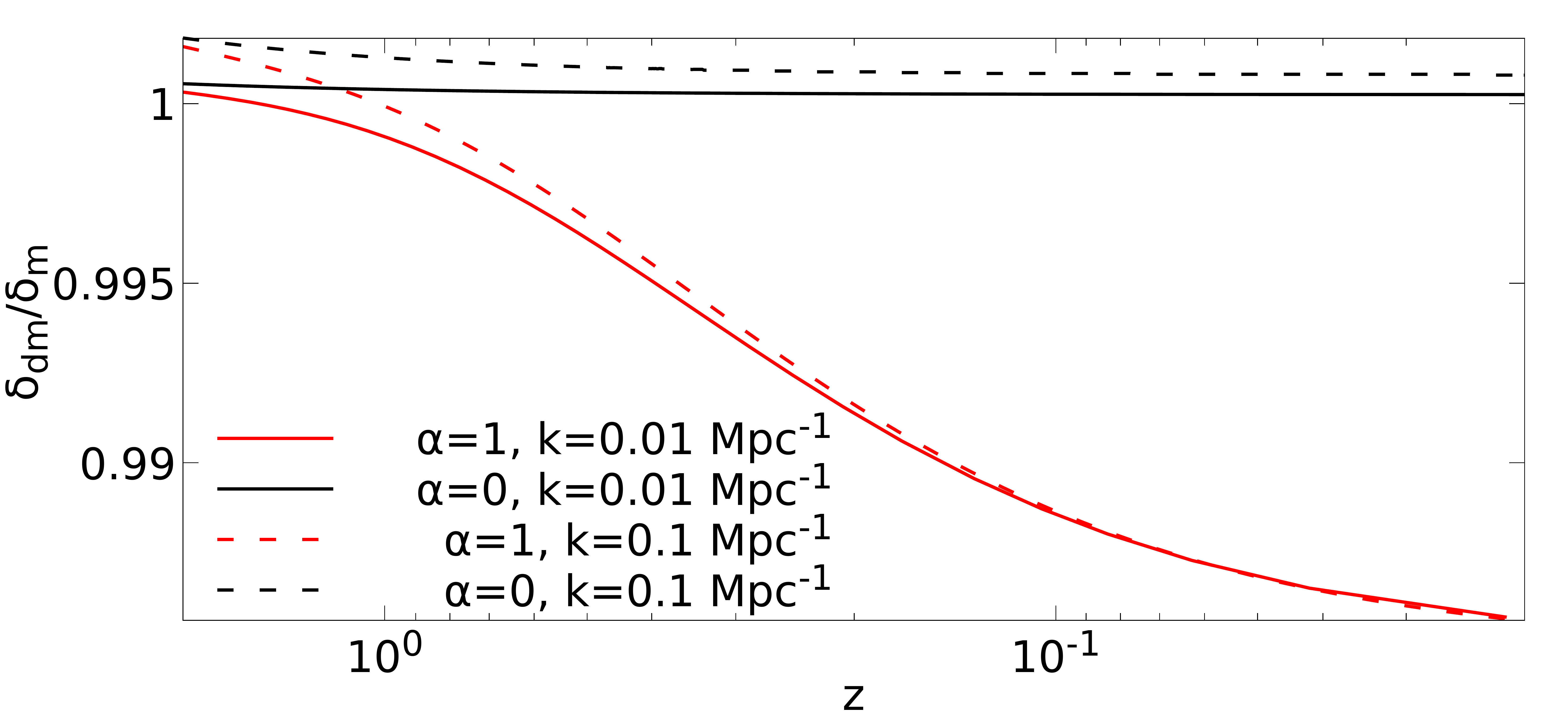}
\caption{In this Figure we plot the ratio of the density contrasts of the dark matter component and the total matter (dark matter plus baryons). We can see that the error of assuming that the ratio $\delta_{\rm dm}/\delta$ is constant and scale independent is below of order 1\%, which, as we will see, is below the sensitivity for the measurement of the interaction parameter $\alpha$. This justifies our assumption since including this effect will lead to a correction smaller than the precision of the measurement.
}
\label{Fig:deltaMdeltaDM}
\end{figure}

The galaxy number counts, see eq.~\eqref{eq:numbercounts}, is sensitive to the radial galaxy velocity. In this section, we consider the assumption that galaxies lie at the bottom of the gravitational potential generated by dark matter halos. From this viewpoint, galaxies are forced to be comoving with the dark matter halos. Therefore, we will have $\ndv^{\rm gal}= \ndv^{DM}$ and we can use the dark matter Euler equation~\eqref{eq:Eulerradial_theta} to 
obtain the relativistic number counts as follows
\be \label{eq:6}
\Delta \left( \bx \right) = b \delta + \HH^{-1} \partial_r \ndv 
+ \left(\frac{\dot\HH}{\HH^2} + \frac{2 - 5s }{r \HH} + 5 s - f_{\rm evo} + \Theta \right) \ndv + \mathcal{O} \left( \HH^2/k^2 \right) \, .
\ee
In this expression, the density contrast is assumed to be evaluated at some redshift $z$, although we will omit the redshift dependence to alleviate the notation. By using the dark matter continuity equation ($\mu = \hat \bx \cdot \hat\bk $)
\be \label{eq7}
\partial_r \ndv^{DM} = f_{DM} \HH \mu^2 \delta_{DM} \qquad \text{and} \qquad  \ndv^{DM}=  { - } i \mu f_{DM} \delta_{DM} \HH/k
\ee
the galaxy number counts becomes proportional to the dark matter density contrast\footnote{In this section we simply denote $f=f_{DM}$ to alleviate the notation.}
\be \label{eq:Delta_Fourierp}
\Delta \left( \bk \right) = b\delta(\bk)+\left[f \mu^2 { - } i \mu  f \frac{\HH}{k}\R { + \left(\frac{\HH}{k} \right)^2\left( \mathcal{F}_0 + \mathcal{F}_2 \mu^2 \right) } \right] \delta_{\rm dm} \left( \bk \right) 
\ee
where 
\be
\label{def:R}
\mathcal{R} = \frac{\dot\HH}{\HH^2} + \frac{2 - 5s }{r \HH} + 5 s - f_{\rm evo} + \Theta 
\ee
and we have introduced $\mathcal{F}_0$ and $\mathcal{F}_2$ to keep track of the subleading relativistic effects, that we will include for consistency in the derivation of the variance. A difficulty arises here because we have two stochastic density variables in~\eqref{eq:Delta_Fourierp}. We will deal with this issue by assuming that the ratio $\delta_{\rm dm}/\delta$ is constant for the scales and redshifts of interest (see Fig.~\ref{Fig:deltaMdeltaDM}). Under this assumption, we can absorb this factor into the growth rate of the interacting dark matter\footnote{And $\mathcal{F}_0$ and $\mathcal{F}_2$, although this will not be relevant since these will not contribute to the dipole covariance at leading order as we will see below.} and write
\be \label{eq:Delta_Fourier}
\Delta \left( \bk \right) = \left[b+f \mu^2 { - } i \mu  f \frac{\HH}{k}\R { + \left(\frac{\HH}{k} \right)^2\left( \mathcal{F}_0 + \mathcal{F}_2 \mu^2 \right) } \right] \delta \left( \bk \right). 
\ee

\subsection{Gravitational potential tracers}
\label{sec:test_vel}

Alternatively, we could consider galaxies as test particles in an external gravitational field where their motion is determined uniquely by the geodesic equation. In this scenario, the motion of galaxies is fully determined by the gravitational potential which in turn is determined by the total matter perturbation via the Poisson equation, which means that we should use the total density contrast in our expressions relating the peculiar velocities with the density perturbation. The two descriptions agree in $\Lambda$CDM, where the geodesic equation reduces to the Euler equation in the matter domination era, but they differ in more generic cases. As shown in Appendix~\ref{app:geodesic} in detail, in this case galaxy geodesic equation leads to the usual Euler equation (i.e.~the Euler equation for non-interacting dark matter) describing the evolution of galaxy velocities~\eqref{eq:Euler}. Therefore, under this assumption the dipole is sensitive to $\Theta$ only 
through its impact on the other cosmological parameters, e.g.~Hubble parameter $\HH$, the amplitude of perturbations $\sigma_8$ or the growth rate $f$. In our case, as the interaction is unable to modify the background cosmology the impact is carried by the growth of structures related variables, like the  growth rate $f$, the gravitational potential $\Phi$ and the $\sigma_8$ parameter. We can therefore simply use eqs.~(\ref{def:R}-\ref{eq:Delta_Fourier}) by setting $\Theta=0$ and considering the growth rate $f$ including baryons.
As we will see in Sec.~\ref{sec:results}, this will reduce the sensitivity of the dipole to $\Theta$, leading to much weaker constraints.

\section{Signature in the dipole of the power spectrum}
\label{sec:dipole}

In order to obtain a first forecast on the ability of upcoming galaxy surveys to constrain or rule out the elastic interacting model, we work in flat-sky approximation\footnote{The amplitude of wide-angle corrections can be reduced by working with a symmetric estimator with respect to the line of sight, see e.g.~Ref.~\cite{Gaztanaga:2015jrs,Castorina:2017inr}.} and neglect integrated effects.\footnote{Since most of the information is carried by the lowest redshift bins, integrated effects have been shown to be small, see Ref.~\cite{Castorina:2021xzs}.} In the flat-sky limit, we can define the dipole estimator as follows
\be \label{eq:dipole_estimator}
\hat P_1^{AB} \left( k \right) = \frac{3}{2 V} \int \frac{\dd\Omega_{\hat \bk}}{2 \pi} \mu \Delta^A \left( \bk \right) \Delta^B \left( -\bk \right)
\ee
where $\mu = \hat \bx \cdot \hat \bk=  - \bn \cdot \hat\bk$ and $V$ is the survey volume.
By computing the power spectrum from eq.~\eqref{eq:Delta_Fourier} and expanding it in multipoles with respect to $\mu$ we find a non-vanishing dipole sourced by relativistic effects for the cross-correlation of two different galaxy populations
\be
P_1^{AB} \left( k \right) = { -} i f \frac{\HH}{k} \left[
\frac{3}{5}  f (\R_A-\R_B)+(b_B \R_A-b_A\R_B)
\right] P \left( k \right)
+   \mathcal{O} \left( \HH^3/k^3 \right) \, .
\ee
In the limit $\mathcal{R}_A = \mathcal{R}_B$ the dipole of the galaxy power spectrum reduces to
\be
P_1^{AB} \left( k \right) = i f \frac{\HH}{k} \left( b_A - b_B\right) \left( \frac{\dot\HH}{\HH^2} + \frac{2 - 5s }{r \HH} + 5 s - f_{\rm evo} + \Theta \right) P\left( k \right) +   \mathcal{O} \left( \HH^3/k^3 \right) \, .
\ee
This equation highlights the importance of the dipole in constraining or detecting the parameter $\Theta$. Indeed, while in other multipoles the elastic scattering between dark matter and dark energy can be detected through the changes in the growth rate and the matter power spectrum only, the dipole depends explicitly on the $\Theta$ parameter.
In order to forecast the detectability of the elastic interaction with future surveys, we need to compute the variance of the dipole (see Appendix~\ref{sec:dipole_covariance} for its derivation)
\bea
&&\hspace{-1cm}
\langle \hat P_1^{AB} \left( k \right) \hat P^{*AB}_1\left(q \right) \rangle_{c} = 
\nonumber \\
&=& \bigg\{
-{ \frac{9}{5}} \left( P_1^{AB} \left( k \right) \right)^2 - { \frac{23}{35}} \left( P_3^{AB} \left( k \right) \right)^2 - { \frac{36}{35}} P_1^{AB} \left( k \right)  P_3^{AB} \left( k \right) 
\nonumber \\
&&
+ 3 \left[  \left( \frac{1}{2} P^{BB}_0 \left( k \right) + \frac{1}{5}P_2^{BB} \left( k \right) \right) N_A + \left(\frac{1}{2} P^{AA}_0 \left( k \right) + \frac{1}{5}P_2^{AA} \right) N_B 
 \right.
 \nonumber \\
 && 
 \left.
 \qquad
+ \frac{1}{2} N_A N_B \right] \bigg\}
\frac{\left( 2 \pi \right)^2}{ V} \frac{\delta_D \left( k - q\right) }{ k^2}
\nonumber \\
&=& \sigma^2_{P_1}\left( k \right)  \frac{\left( 2 \pi \right)^2}{ V} \frac{\delta_D \left( k - q\right) }{ k^2}
\, . 
\label{eq:covariance_dipole}
\eea
From this variance, we clearly see the importance of the dipole in detecting relativistic effects and, therefore, deviations from the Euler equation induced by the model under consideration. Indeed, while the dipole is suppressed by a factor $\HH/k$ with respect to the even multipoles, see eqs.~(\ref{eq:P0}-\ref{eq:P4}), its variance is not affected by the cosmic variance of the even multipoles. Therefore the detection of the dipole is mostly limited by shot-noise, as we will confirm with our numerical results in the next section.

\section{Numerical results}
\label{sec:results}
In this section we study the signal-to-noise ratio to detect a non-vanishing value for the parameter $\Theta$ through the dipole of the galaxy power spectrum. In our case, a non-zero value of $\Theta$ is associated to the momentum transfer between dark energy and dark matter that modifies the Euler equation as explained in Section~\ref{sec:deviation_Euler}. Such momentum transfer is controlled by the coupling parameter of the model $\alpha$, as summarised in Section~\ref{sec:model}. For this purpose, we compute the Fisher element $F_{\alpha \alpha}$ defined as 
\be
F_{\alpha \alpha} = \sum_i \frac{V_i}{4 \pi^2}\int \dd k k^2  \left| \frac{\partial P_1 \left( k ,z_i\right)}{\partial \alpha} \right|^2 \sigma^{-2}_{P_1} \left( k,z_i \right) \;,
\label{eq:Fisher_element}
\ee 
where the coupling parameter of the model $\alpha$ is related to $\Theta$ as follows
\be \label{eq:Theta_alpha}
\Theta=\frac{H_0}{\Omega_{DM}(z)\HH(z)(1+z)}\alpha\;.
\ee
Since $\alpha$ has no impact on the background cosmological quantities like $\mathcal{H}$, $H_0$ and $\Omega_{\rm i}$, the Fisher elements of $\alpha$ and $\Theta$ are trivially related. Then, as we are only varying one parameter we obtain the expected uncertainty from the Fisher element simply as
\be
\Delta\alpha = \frac{2}{F_{\alpha \alpha}^{1/2} }\;.
\ee
We obtain the different cosmological functions such as the matter power spectrum or the Hubble function from a modified version of the publicly available CLASS code~\cite{2011JCAP...07..034B} developed in Ref.~\cite{Figueruelo:2021elm}. This modified code accounts for the effects of the interaction as it includes the new terms that appear in the Euler equations due the  dark matter-dark energy momentum transfer [see eqs.~\eqref{eq:thetaDM_alpha} and~\eqref{eq:thetaDE_alpha}].  We perform our calculation using as background cosmology a $w$CDM model since the interaction necessitates $w\neq-1$, as required by the Euler equations~\eqref{eq:thetaDM_alpha} and~\eqref{eq:thetaDE_alpha}. We compute the quantities related to the perturbation sector using  the Newtonian gauge. We set the coupling parameter to $\alpha=1$ for our fiducial cosmology motivated by the best fit value to cosmological observations of the model~\cite{Asghari:2019qld,Figueruelo:2021elm}. We remind the reader that the parameter $\alpha$ is the only new parameter in this model, which controls the efficiency of the momentum transfer between dark matter and dark energy. The remaining parameters of our fiducial cosmology are set to $ \Omega_{\rm b}h^2 = 0.02264$, $\Omega_{\rm c} = 0.1163$, $n_s = 0.9721$, $A_s = 2.063\;10^{-9}$, $\tau =  0.0502$, $w = -0.948$ and $h = 0.6788$, corresponding to the results obtained in Ref.~\cite{Figueruelo:2021elm} using cosmological data of Planck 2018 TT, TE, EE and lensing data~\cite{PLANCK2018,Planck2019V} + JLA~\cite{JLA} + BAO~\cite{BAO1,BAO2,BAO3} + PlanckSZ~\cite{PlanckSZ} + CFHTLens~\cite{CFHTL} (see the full explanation of Section 4 of~\cite{Figueruelo:2021elm} for more details). In any case, our results are not expected to have a strong  dependence on the fiducial cosmology used (with the exception of the value of $\alpha$ of course). Background and early Universe parameters or quantities are insensitive to the interaction by the very intrinsic nature of the coupling, but the only strong correlation is with the value of $\sigma_8$. Since we use $\sigma_8$ as a derived parameter, then we do not fix its value, the dependence with the fiducial cosmology is mainly encoded in the value of $\alpha$ for a reasonable choice of parameters.

In the following we will analyse the two scenarios explained above for the velocity of galaxies, namely: galaxies as tracers of the dark matter velocity field and galaxies as tracers of the gravitational potential. We will assume that the two populations of objects are split evenly, that is the density  of each population fulfills $n_A \left( z \right)=n_B \left( z \right) = n \left( z \right)/2$ where $n \left( z \right)$ is the value reported in  Table~\ref{tab:densities_bias}. For the bias, we will follow the prescription 
\bea
b_A\left( z \right) &=& b \left( z \right) + 0.25\;, \\
b_B\left( z \right) &=& b \left( z \right) - 0.25\;,
\eea
such that $b_A\left( z \right)- b_B\left( z \right)=0.5$ as reported in Ref.~\cite{Bonvin:2018ckp}. We will also consider the configuration with $f_{\rm evo} = 0$\footnote{In the next paragraph we study what happens when $f_{\rm evo} \neq 0$.}  for both $A$ and $B$ populations and we also set $s_A = s_B = 0$. Considering other scenarios where $f_{{ \rm evo},A}\neq f_{{\rm evo},B}$ and/or $s_A \neq s_B $ will only improve our results, since in general the more pronounced the differences between both populations the larger the detection capabilities. Thus, our analysis will be conservative.
For the shot-noise, we consider the simple prescription $N_A=1/n_A$ and $N_B=1/n_B$ while the volume is defined as 
\be
V_{i} = \frac{4 \pi}{3}f_{\rm sky} \Big[ r \left( z_{\rm max}\right)^3 -r\left(z_{\rm min}\right)^3    \Big]\;,
\ee
where $r(z)$ is the comoving distance to redshift $z$ and $f_{\rm sky}$ is the fraction of the sky, that for the case of Square
Kilometre Array (SKA)~\cite{SKA:2018ckk} survey is $f_{\rm sky} = 30000/(360^2/\pi)\simeq 0.73$.

\begin{table}[]
\renewcommand{\arraystretch}{1.2} 
\centering
\begin{tabular}{|c|c||c|c|}
\hline
$z_{min}$ & $z_{min}$ & $n(z)\; [Mpc^{-3}]$ & $b(z)$  \\ \hline\hline
$0.1$ & $0.2$ & $6.20\;10^{-2} $ & $0.623$   \\ \hline
$0.2$ & $0.3$ & $3.63\;10^{-2} $ & $0.674$   \\ \hline
$0.3$ & $0.4$ & $2.16\;10^{-2} $ & $0.730$    \\ \hline
$0.4$ & $0.5$ & $1.31\;10^{-2} $ & $0.790$    \\ \hline
$0.5$ & $0.6$ & $8.07\;10^{-3} $ & $0.854$   \\ \hline
$0.6$ & $0.7$ & $5.11\;10^{-3} $ & $0.922$    \\ \hline
$0.7$ & $0.8$ & $3.27\;10^{-3} $ & $0.996$    \\ \hline
$0.8$ & $0.9$ & $2.11\;10^{-3} $ & $1.076$   \\ \hline
$0.9$ & $1.0$ & $1.36\;10^{-3} $ & $1.163$    \\ \hline
$1.0$ & $1.1$ & $8.70\;10^{-4} $ & $1.257$   \\ \hline
$1.1$ & $1.2$ & $5.56\;10^{-4} $ & $1.360$   \\ \hline
$1.2$ & $1.3$ & $3.53\;10^{-4} $ & $1.472$   \\ \hline
$1.3$ & $1.4$ & $2.22\;10^{-4} $ & $1.594$   \\ \hline
$1.4$ & $1.5$ & $1.39\;10^{-4} $ & $1.726$    \\ \hline
$1.5$ & $1.6$ & $8.55\;10^{-5} $ & $1.870$    \\ \hline
$1.6$ & $1.7$ & $5.20\;10^{-5} $ & $2.027$   \\ \hline
$1.7$ & $1.8$ & $3.12\;10^{-5} $ & $2.198$   \\ \hline
$1.8$ & $1.9$ & $1.83\;10^{-5} $ & $2.385$   \\ \hline
$1.9$ & $2.0$ & $1.05\;10^{-5} $ & $2.588$   \\ \hline
\end{tabular}
\caption{We consider the galaxy bias $b\left( z \right)$ and the number density as reported in Table~3 of Ref.~\cite{Bull:2015lja} for SKA.}
\label{tab:densities_bias}
\end{table}

\subsection{Galaxies as dark matter tracers}
In the first scenario galaxies are assumed to be perfect tracers of the dark matter velocity field. Under this assumptions and using a fiducial cosmology with~$\alpha=1$, as motivated by the fits to current data, we obtain the following expected uncertainty from the measurement of the dipole:
\be
 \Delta_{1\sigma}\alpha=0.198\;.
 \label{eq:DeltaalphaDM}
\ee
This is a main result of this paper, namely: with future SKA-like surveys we should definitely see the interaction. This result represents a clear prediction of the model and, therefore, it provides a smoking gun for these models. Let us recall that, although we are focusing on the covariantised Thomson-like scattering in the present work, this is utilized only as a proxy for the general class of models with pure momentum exchange so we expect similar results for those models within this class able to alleviate the $\sigma_8$ tension. Furthermore, the latest data available gives a value for the coupling parameter ${\alpha=1.01_{-0.33}^{+0.26}}$~\cite{Figueruelo:2021elm} so, with the employed specifications, we will be able to improve the best current constraints on the interaction parameter. 

For completeness, we will explore the impact of having different evolution bias for both populations. A simple prescription would be $f_{\rm evo}= (b-1) f \delta_c$~\cite{Jeong:2011as} given the different bias $b$ of population A and population B as $b_A\left( z \right)- b_B\left( z \right)=0.5$, where\footnote{We are taking the \LCDM{} value of $\delta_c$. In our scenario, this value will receive corrections from the effects of the interaction in the clustering. Since our interaction is relevant only at low redshift the spherical collapse model remains unaffected for most of the structure formation. A full account of the effect of the interaction on the spherical collapse and the value of $\delta_c$ (that presumably will even become scale-dependent) will require a detailed treatment. We do not expect however a large deviation from the standard model so the corrections to our results are not  expected to be relevant. In any case, the corrections will not spoil our main result about the detectability of the interaction.} $\delta_c \simeq 1.686$. For simplicity we keep the magnification bias as before. Then, we obtain $\Delta\alpha=0.166$. As expected, the more the differences between both populations the better the constraining power. Although one could wonder the effect of including a different magnification bias for each population $s_A\neq s_B$ we have seen the improvement is modest when changing the evolution bias. 
Moreover, as can be observed from Fig.~\ref{Fig:alpha_fevo}, most of the constraining power comes from the lowest redshift bins, where galaxy surveys are more complete and so the magnification bias becomes closer to $0$. However, we should not forget that, even in the previous scenario where both evolution and magnification bias are the same for both populations, an SKA-like experiment would have for a fiducial cosmology $\alpha=1$ a predicted uncertainty of $\Delta\alpha=0.198$, which will improve the constraints obtained with the latest experiments as obtained in Refs.~\cite{Asghari:2019qld,Figueruelo:2021elm}.

\subsection{Galaxies as tracers of the gravitational potential}
In the scenario where galaxies are just test particles moving within the gravitational potential, we obtain the forecasted uncertainty
\be
\Delta_{1\sigma}\alpha=1.30\;,
\ee
which is substantially worse than~\eqref{eq:DeltaalphaDM}. In particular, this uncertainty in the interaction parameter $\alpha$ means that, under these circumstances, the dipole measurements will not show a clear signal of the interaction. The reason for the worsening of the result is that now the motion of galaxies is not directly affected by the interaction, since their cosmological evolution is determined by an unmodified Euler equation, but only indirectly through the effects on the gravitational potential, which is modified by the different clustering of the dark matter. This induces a decoupling between the motion of galaxies and dark matter. In other words galaxies are no longer a good tracer of the dark matter velocity field where the interaction plays a predominant role at low redshift. As commented above, we believe the scenario with galaxies being good tracers of the dark matter velocity field is more realistic, so the actual situation would yield the tighter constraint~\eqref{eq:DeltaalphaDM}.

\begin{figure}[ht!]
\centering
\includegraphics[scale=0.23]{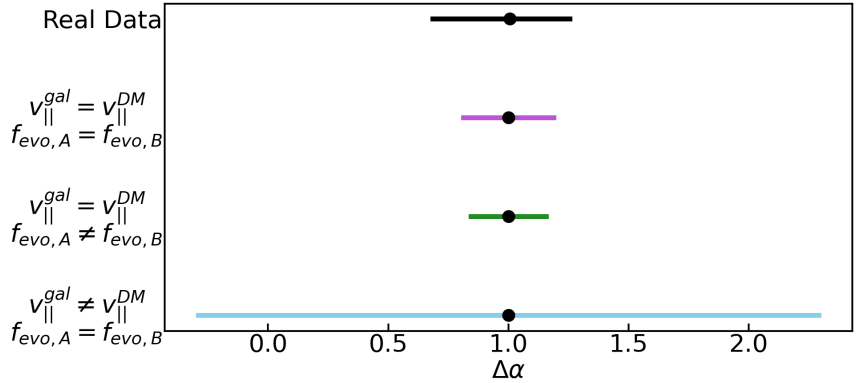}
\caption{Constraining power for the case when dark matter halos and galaxies are comoving with $f_{\rm evo,A}=f_{\rm evo,B}=0$ (purple), with $f_{\rm evo,I}= (b_I-1) f \delta_c $, $I=A,B$ (green) and the case when galaxies are just test particles moving with the gravitational potential created by dark matter halos (blue). The black line corresponds to the constraints obtained with real data for this mode in Ref.~\cite{Figueruelo:2021elm}. 
}
\label{Fig:Result_compared}
\end{figure}

\begin{figure}[ht!]
\centering
\includegraphics[scale=0.15]{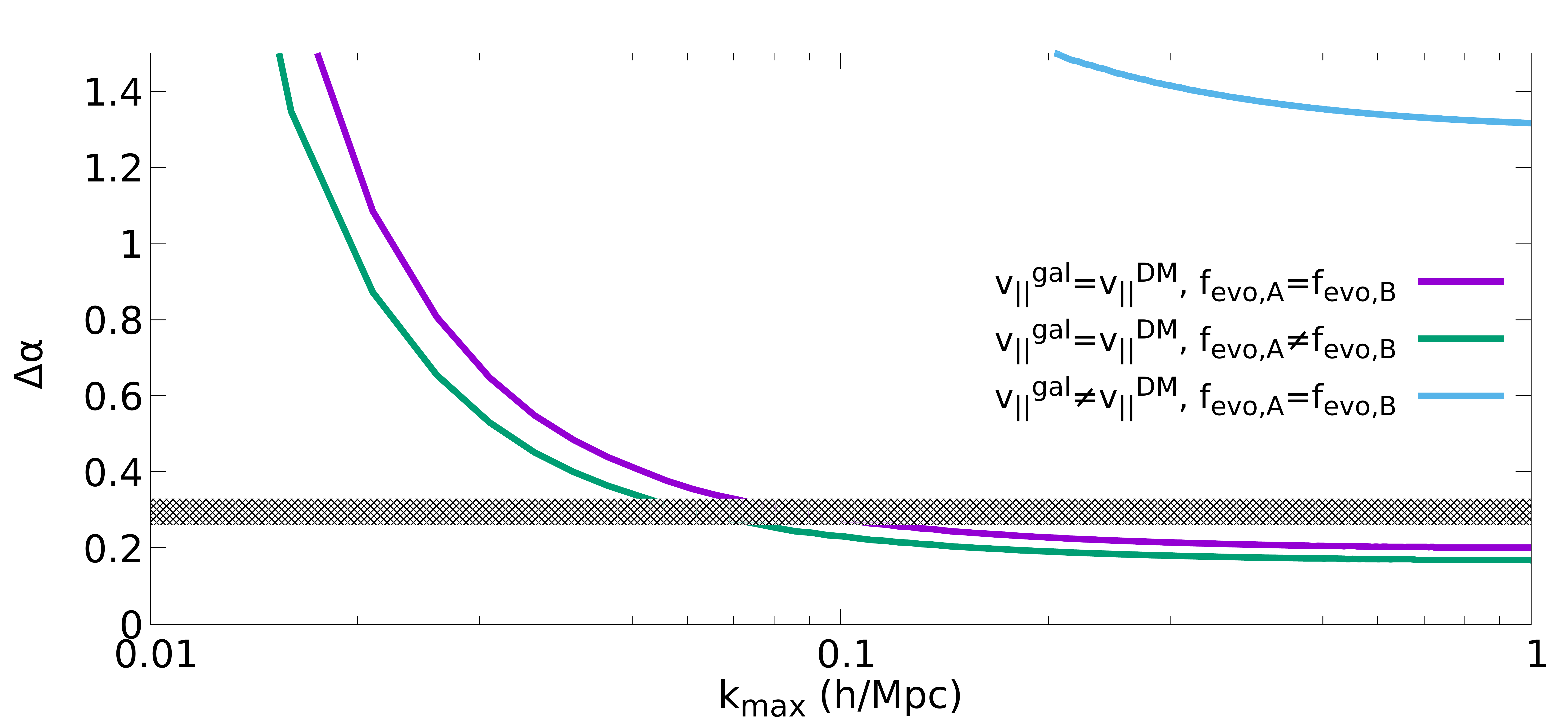}
\caption{Constraining power as a function of $k_{\rm max}$ used for the case when dark matter halos and galaxies are comoving with $f_{\rm evo,A}=f_{\rm evo,B}=0$ (purple), with $f_{\rm evo,I}= (b_I-1) f \delta_c $, $I=A,B$ (green) and the case when galaxies are just test particles moving with the gravitational potential created by dark matter halos (blue). The black region corresponds to the constraints obtained with real data for this mode in Ref.~\cite{Figueruelo:2021elm}.}
\label{Fig:Result_as_k_max}
\end{figure}

To conclude, we can see in Figure~\ref{Fig:Result_compared} the expected constraints for the model parameter $\alpha$ considering  the case where galaxies are comoving with dark matter halos both with $f_{\rm evo,I}= 0$ (purple line) and with $f_{\rm evo,I}= (b_I-1) f \delta_c $ (green line), and the scenario where galaxies are just test particles (blue line). In black we have the result obtained in Ref.~\cite{Figueruelo:2021elm} where they use the last available data to constrain the parameter obtaining ${\alpha=1.01_{-0.33}^{+0.26}}$. The same is depicted in Figure~\ref{Fig:Result_as_k_max} as function of the maximum scale $k_{\rm max}$ used in the computation of the Fisher element [see equation~\eqref{eq:Fisher_element}] of the coupling $\alpha$. Therefore,  we can conclude that as long as galaxies are comoving with dark matter halos, future experiments will have the potential to improve the constraints of the model parameter $\alpha$, compared to current experiments. However, if the comoving assumption is not valid, an SKA-like experiment will not be able to improve our current results as the effect will be diluted since the interacting partner is not baryons, that is galaxies, but dark matter.

From Figure~\ref{Fig:Result_as_k_max}, we observe that most of the information is obtained from the linear regime, in particular from scales below $k_{\rm max}< 0.1 h/{\rm Mpc}$.
As commented in the previous section, regardless of the suppression factor $\HH/k$ in the dipole, we do not obtain a relevant amount of physical information from scales comparable to the horizon, i.e.~$k\sim \HH$, and the detectability of the dipole, and therefore of $\alpha$, depends on the bias difference of the two populations and is limited by shot-noise on small scales.

\section{Conclusions}
\label{sec:conclusions}
In this work we have explored the ability of future SKA-like measurements of the dipole of the galaxy power spectrum to detect a class of interactions between dark matter and dark energy with pure momentum exchange introduced in~\cite{Asghari:2019qld,Figueruelo:2021elm}. In previous works~\cite{Bonvin:2018ckp,Bonvin:2020cxp}, it was established that certain corrections to the dark matter Euler equation could be detected with the dipole of the galaxy power spectrum. We have shown how these corrections naturally arise within the considered interacting models via the momentum exchange between dark matter and dark energy and exploited this fact to show that the interaction affects the dipole of the galaxy power spectrum.

Current observations spark a debate on the detection of this class of  interactions and the role of low-$z$ surveys. When using such low-$z$ data-sets like Planck Sunyaev-Zeldovich~\cite{PlanckSZ} information from cluster counts, the interacting model is clearly favoured over \LCDM{}, with a measurement of the interaction at more than 3$\sigma$ significance. When only using Planck data or with Supernovae and/or BAO data such detection is unattainable and the model performance is similar to \LCDM{} with no statistically significant difference. This feature is not exclusive of the model studied here, but it emerges as a generic feature of momentum transfer for a variety of realisations~\cite{Pourtsidou:2016ico,Lesgourgues:2015wza,Linton:2021cgd,Jimenez:2020ysu} (see also~\cite{Figueruelo:2021elm,BeltranJimenez:2021wbq} for detailed discussion). Thus, we interpret the considered model as a proxy and our findings are expected to apply to similar scenarios where a momentum exchange of dark matter gives appreciable effects, in particular those able to alleviate the $\sigma_8$ tension. Thus, since fits of the interacting model to some sets of current data favour a detection of the interaction with a definite prediction for the strength of the interaction, the effect on the dipole provides a smoking gun for these models. Our goal  in this work has been to determine weather future measurements of the dipole could indeed observe these effects.

A cautionary remark should be made because the interaction only affects the dark matter component, while baryons evolve as usual, so some care need to be taken in the computation of the relevant velocities and growth factors. This can be regarded as a sort of fake violation of the equivalence principle since dark matter and baryons evolve differently. Of course, this is simply due to the extra friction that dark matter suffers because of its interaction with dark energy. To tackle this issue, we have employed two scenarios: One with galaxies being perfect tracers of the dark matter velocity field, and another one where galaxies are detached from dark matter halos and move as test particles. We have obtained that the first scenario clearly permits to observe the effects of the interaction on the dipole, even improving current constraints, while the second scenario is not sufficiently sensitive to observe them.

The question is then which scenario is more realistic. We have argued in this work that the first one with galaxies providing good tracers of the underlying dark matter velocity field seems more realistic so we would be in the optimal scenario to see the interaction in the dipole. Although this might seem wishful-thinking to some extend, we should notice that galaxies are virialised objects within dark matter halos and the preferred value for the interaction is not large so we find it unlikely that it will drag galaxies away from their host dark matter haloes.\footnote{This applies to dark matter dominated objects. The situation would be different for objects with a low concentration of dark matter.} In order to clarify the viability of this scenario, numerical simulations will be necessary. We are currently investigating this issue~\cite{SimulationsDMDEinteraction}. Numerical simulations for an analogous model featuring elastic interactions in the dark sector were performed~\cite{Baldi:2014ica,Baldi:2016zom}, but they did not consider a distinction between baryons and dark matter particles. In~\cite{Ferlito:2022mok} the authors carried out numerical simulations for a model featuring an elastic interaction between dark energy and baryons, which is radically different from the models under consideration in this work so their results cannot be used to decide between our two scenarios. The results of this work could however be used for the analogous model explored in~\cite{BeltranJimenez:2020qdu}.

In summary, we can conclude that upcoming galaxy surveys will certainly have the statistical power to detect the possible signature of the elastic interactions in the dark sector under conservative conditions. If confirmed, our results will promote the dipole of the galaxy power spectrum to be a clear smoking gun to robustly corroborate the evidence in favour of the momentum exchange in the dark sector, or else to completely rule it out. In any case, we believe our findings further calls for a deeper scrutiny of this class of models.

\acknowledgments
We thank Dario Bettoni, Pierre Fleury and Florencia A. Teppa Pannia for useful discussions. This work has been funded by Project PID2021-122938NB-I00 funded by the Spanish ``Ministerio de Ciencia e Innovaci\'on" and FEDER ``A way of making Europe". DF acknowledges support from the programme {\it Ayudas para Financiar la Contrataci\'on Predoctoral de Personal Investigador (ORDEN EDU/601/2020)} funded by Junta de Castilla y Le\'on and European Social Fund.

\appendix

\section{Dipole Variance}
\label{sec:dipole_covariance}
In this section we compute the variance of the dipole defined in eq.~\eqref{eq:dipole_estimator}, starting from
\bea
&&\langle \hat P_1^{AB} \left( k \right) \hat P^{*AB}_1\left(q \right) \rangle_{c} =\frac{9}{4 V^2} \int \frac{\dd\Omega_{\hat \bk}}{2 \pi} \frac{\dd \Omega_{\hat \bq}}{2 \pi} \mu_k \mu_q \langle 
\Delta^A \left( \bk \right) \Delta^B \left(- \bk \right) \Delta^A \left( -\bq \right) \Delta^B \left( \bq \right)  
\rangle
\nonumber \\
&=&\frac{9}{4 V^2} \delta_D^{(3)} \! \left( 0 \right) \left( 2 \pi \right)^6  \int \frac{\dd\Omega_{\hat \bk}}{2 \pi} \frac{\dd \Omega_{\hat \bq}}{2 \pi} \mu_k \mu_q \left[ P^{AA} \left( \bk \right)  P^{BB} \left( - \bk \right) \delta_D \left( \bk - \bq  \right)
\right.
\nonumber \\
&&
\left.
\qquad \qquad \qquad \qquad 
+ P^{AB} \left( \bk \right)  P^{BA} \left( - \bk \right) \delta_D \left( \bk + \bq  \right)
\right]
\nonumber \\
&=&
\frac{9}{4} \frac{\left( 2 \pi \right)^6}{V^2} \frac{\delta_D^{(3)} \! \left( 0 \right)}{ 2 \pi} \frac{\delta_D \left( k - q \right) }{k^2}
\nonumber \\
&&
\times\sum_{\ell_1 \ell_2} \Big[   P^{AA}_{\ell_1}\left( k \right)  P^{BB}_{\ell_2}\left( k \right)
- P^{AB}_{\ell_1}\left( k \right)  P^{BA}_{\ell_2}\left( k \right)
\Big]
\int \dd\mu \mu^2 \mathcal{L}_{\ell_1} \left( \mu \right)  \mathcal{L}_{\ell_2} \left(- \mu \right) \, .
\label{eq:2}
\eea
Before proceeding to simplifying the variance, we will compute the multipole expansion of the power spectrum
\bea
P^{AB} \left( \bk \right) \;=&& \left[ b_A + f \mu^2 { -} i \mu f \frac{\HH}{k} \mathcal{R}_A + { \left( \mathcal{F}_0^A +\mathcal{F}_2^A \mu^2 \right) \left( \frac{\HH}{k}\right)^2  } \right] 
\nonumber \\
&\times&\left[ b_B + f \mu^2 { +} i \mu f \frac{\HH}{k} \mathcal{R}_B + { \left( \mathcal{F}_0^B +\mathcal{F}_2^B \mu^2 \right) \left( \frac{\HH}{k}\right)^2  } \right]  P\left( k \right) \, .
\eea
By decomposing into Legendre polynomials, we obtain the following multipole coefficients:
\bea
\label{eq:P0}
P_0^{AB} \left( k \right) &=&\left[  b_A b_B  + \frac{1}{3} \left( b_A + b_B \right) f 
\right.
\nonumber \\
&& \left.
+\frac{f^2}{5}  
\left( \frac{1}{3} (b_A (3 \mathcal{F}_0^B+\mathcal{F}_2^B)+b_B (3 \mathcal{F}_0^A+\mathcal{F}_2^A))+\frac{1}{3} f^2 \mathcal{R}_A \mathcal{R}_B
\right. \right.
\nonumber \\
&&
\left. \left.
+\frac{1}{15} f (5 \mathcal{F}_0^A+5 \mathcal{F}_0^B+3 (\mathcal{F}_2^A+\mathcal{F}_2^B)) \right) \left(\frac{\HH}{k} \right)^2
\right] P \left( k \right) 
\nonumber \\
&&
+   \mathcal{O} \left( \HH^3/k^3 \right) 
\, , \\
\label{eq:P1}
P_1^{AB} \left( k \right) &=& { -} i f \frac{\HH}{k} \left[
\frac{3}{5}  f (\R_A-\R_B)+(b_B \R_A-b_A\R_B)
\right] P \left( k \right)
\nonumber \\
&&
+   \mathcal{O} \left( \HH^3/k^3 \right) 
\label{P1}
\, , \\
\label{eq:P2}
P_2^{AB} \left( k \right) &=&  \left[ \frac{2}{3} f (b_A+b_B)
\right.
\nonumber \\
&&
\left.
+\frac{4 f^2}{7}   
\left(
\frac{2}{3} (b_A \mathcal{F}_2^B+b_B \mathcal{F}_2^A)+\frac{2}{3} f^2 \mathcal{R}_A \mathcal{R}_B
\right. \right.
\nonumber \\
&&
\left. \left.
+\frac{2}{21} f (7 \mathcal{F}_0^A+7 \mathcal{F}_0^B+6 (\mathcal{F}_2^A+\mathcal{F}_2^B)) \right) \left( \frac{\HH}{k} \right)^2
\right] P \left( k \right)
\nonumber \\
&&
+   \mathcal{O} \left( \HH^3/k^3 \right)
\, , \\
\label{eq:P3}
P_3^{AB} \left( k \right) &=& { -} \frac{2}{5} i f^2 \frac{\HH}{k} \left[ \mathcal{R}_A - \mathcal{R}_B    \right] P \left( k \right)
+   \mathcal{O} \left( \HH^3/k^3 \right)
\, , \\
\label{eq:P4}
P_4^{AB} \left( k \right) &=& \left[ \frac{8}{35} f^2 
+ \left( \frac{8}{35} f (\mathcal{F}_2^A+\mathcal{F}_2^B) \right) \left( \frac{\HH}{k} \right)^2
\right]P \left( k \right) +   \mathcal{O} \left( \HH^3/k^3 \right)
\, .
\eea
As expected the odd multipoles are sourced by the relativistic effects only, while these contribute to the even multipoles only at subleading order $\mathcal{O} \left( \HH^2/k^2 \right)$. Moreover the odd multipoles do not vanish only if we consider two different tracers $A \neq B$.
At this point we can compute the sum in eq.~\eqref{eq:2} as
\bea
&&\hspace{-1.5cm}\langle \hat P_1^{AB} \left( k \right) \hat P^{*AB}_1\left(q \right) \rangle_{c}
= \frac{\delta_D^{(3)}\left( 0 \right) }{ 2 \pi } \frac{\left( 2 \pi \right)^6}{ V^2} \frac{\delta_D \left( k - q\right) }{ k^2}\left[{ \frac{9}{5}} \left(b_B \R_A - b_A \R_B \right)^2
\right.
\nonumber \\
&&
\left.
+ { \frac{18}{7}} f \left( \R_A - \R_B \right)  \left(b_B \R_A - b_A \R_B \right)
+ { f^2} \left( \R_A - \R_B \right) \right]
\frac{\HH^2}{k^2} f^2 P^2 \left( k \right)
\nonumber \\
&=&  \frac{\left( 2 \pi \right)^2}{ V} \frac{\delta_D \left( k - q\right) }{ k^2}\left[ { \frac{9}{5}} \left(b_B \R_A - b_A \R_B \right)^2
\right.
\nonumber \\
&&
\left.
+ { \frac{18}{7}} f \left( \R_A - \R_B \right)  \left(b_B \R_A - b_A \R_B \right)
+ { f^2} \left( \R_A - \R_B \right)\right]
\frac{\HH^2}{k^2} f^2 P^2 \left( k \right) \, .
\eea
where we have used that $\delta_D^{(3)}\left( 0 \right)  \simeq V / \left( 2 \pi \right)^3$ for a finite volume survey. Let us emphasise that the sub-leading terms $\mathcal{F}_0$ and $\mathcal{F}_2$ do not contribute to the dipole covariance at the order $\mathcal{O} \left( \HH^2/k^2 \right)$.

We also need to include the shot-noise contribution. Only the monopole of the auto-correlation power spectra have a non-vanishing shot-noise contribution. Therefore we need to add to eq.~\eqref{eq:2} the following term
\bea
P_{\ell_1}^{AA} \left( k \right) P_{\ell_2}^{BB} \left( k \right) &\rightarrow& 
\left( P_{\ell_1}^{AA} \left( k \right) + \delta_{\ell_1 0 } N_A \right) \left(  P_{\ell_2}^{BB} \left( k \right)+ \delta_{\ell_2 0 } N_B \right) 
\nonumber \\
&=& P_{\ell_1}^{AA} \left( k \right) P_{\ell_2}^{BB} + \delta_{\ell_2 0}P_{\ell_1}^{AA} N_B + \delta_{\ell_1 0}P_{\ell_1}^{BB} N_A + \delta_{\ell_1 0}\delta_{\ell_2 0} N_A N_B\;,
\nonumber \\
&&
\eea
where $N_A$ and $N_B$ are the shot-noise power spectra of the populations $A$ and $B$ respectively. 
Combining all together we find the following variance for the dipole of the power spectrum
\bea
&&\hspace{-1cm}
\langle \hat P_1^{AB} \left( k \right) \hat P^{*AB}_1\left(q \right) \rangle_{c} = 
\nonumber \\
 &=&   \bigg\{\left[ \frac{9}{10} \left(b_B \R_A - b_A \R_B \right)^2+ \frac{9}{7} f \left( \R_A - \R_B \right)  \left(b_B \R_A - b_A \R_B \right)
 \right.
 \nonumber \\
 && 
 \left.
 \qquad
+ \frac{f^2}{2} \left( \R_A - \R_B \right)^2 \right]
\frac{\HH^2}{k^2} f^2 P^2 \left( k \right)  
\nonumber \\
&&+ \left[ \frac{3}{2} \left( b_A^2 {N_B}+ b_B^2 {N_A}\right)+\frac{9}{5} f ( {b_A} {N_B}+ {b_B} {N_A})+\frac{9}{14} f^2 ({N_A}+{N_B})
 \right.
 \nonumber \\
 && 
 \left.
 \qquad
{ + \frac{9}{10} f^2 \frac{\HH^2}{k^2} \left(N_B \mathcal{R}_A^2 + N_A \mathcal{R}_B^2 \right)}
\right] P \left( k \right) 
+ \frac{3}{2} N_A N_B \bigg\} \frac{\left( 2 \pi \right)^2}{ V} \frac{\delta_D \left( k - q\right) }{ k^2}
\nonumber \\
&=& \bigg\{
-{ \frac{9}{5}} \left( P_1^{AB} \left( k \right) \right)^2 - { \frac{23}{35}} \left( P_3^{AB} \left( k \right) \right)^2 - { \frac{36}{35}} P_1^{AB} \left( k \right)  P_3^{AB} \left( k \right) 
\nonumber \\
&&
+ 3 \left[  \left( \frac{1}{2} P^{BB}_0 \left( k \right) + \frac{1}{5}P_2^{BB} \left( k \right) \right) N_A + \left(\frac{1}{2} P^{AA}_0 \left( k \right) + \frac{1}{5}P_2^{AA} \right) N_B 
 \right.
 \nonumber \\
 && 
 \left.
 \qquad
+ \frac{1}{2} N_A N_B \right] \bigg\}
\frac{\left( 2 \pi \right)^2}{ V} \frac{\delta_D \left( k - q\right) }{ k^2}
\nonumber \\
&=& \sigma^2_{P_1}\left( k \right)  \frac{\left( 2 \pi \right)^2}{ V} \frac{\delta_D \left( k - q\right) }{ k^2}
\, . 
\label{sP12}
\eea

\section{Degeneracy with evolution bias}

\begin{figure}[ht!]
\centering
\includegraphics[scale=0.2]{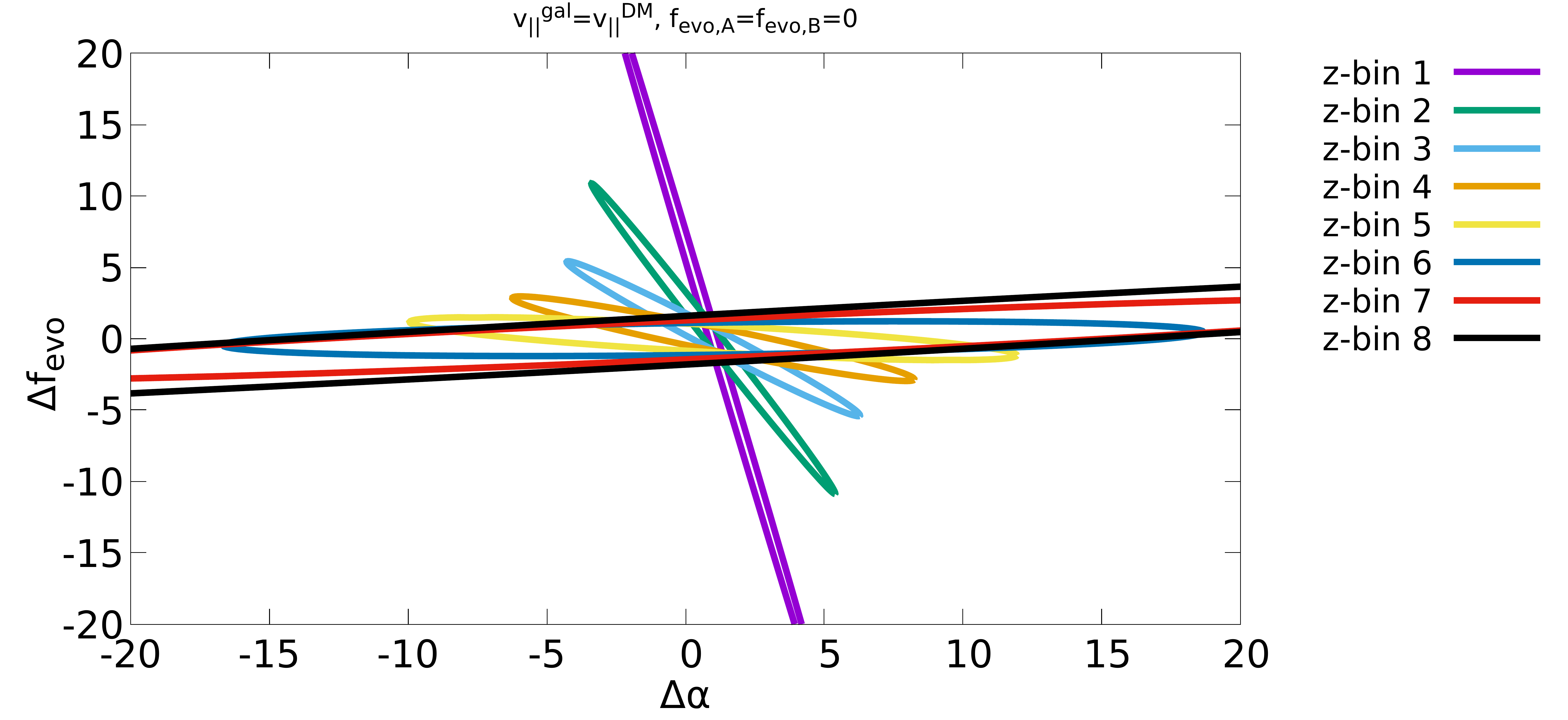}
\includegraphics[scale=0.2]{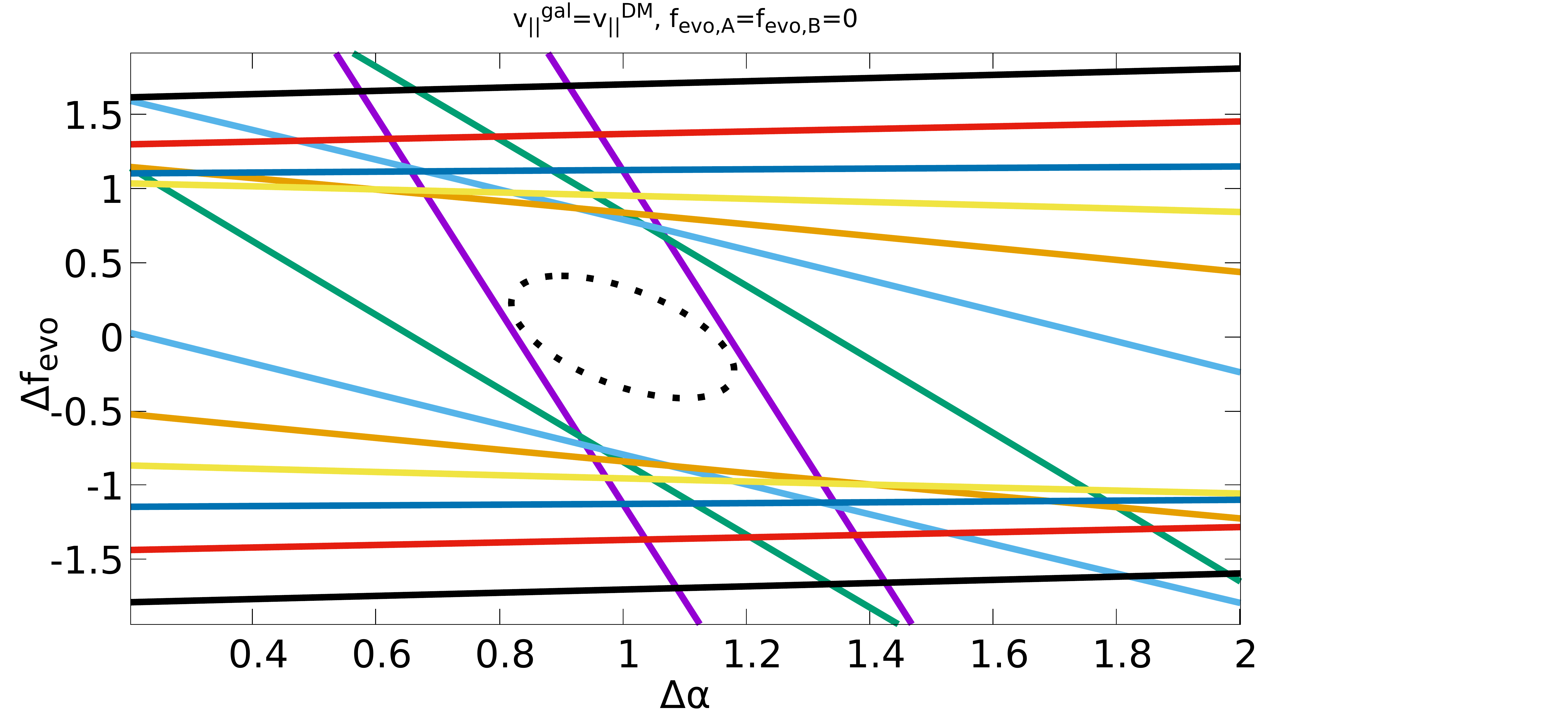}
\caption{We show the 2-dimensional contour plot for the evolution bias $f_{\rm evo}$ and the parameter $\alpha$. Different colors refer to individual redshift bins. In the bottom panel we zoom in to see how the different degeneracies, generated by the redshift evolution of $\alpha$, strongly reduce the error-bars in $f_{\rm evo}$ and $\alpha$. We can also see that most of the information is carried by the lower redshift bins, where the dipole amplitude is larger.}
\label{Fig:alpha_fevo}
\end{figure}

In this Appendix we will comment on the apparent degeneracy that might be expected by looking at eq.~\eqref{eq:6} between $\Theta$ and $f_{\rm evo}$. However, the relation between $\Theta$ and $\alpha$ is redshift-dependent, see eq.~\eqref{eq:Theta_alpha}, and this together with the dependence on the power spectrum and the growth rate on $\alpha$ will break the degeneracy. In the following we will show how well this degeneracy is broken in more detail. We indeed include $f_{\rm evo}$ in our Fisher analysis as 
\begin{equation}
    F=\begin{pmatrix}
     F_{\alpha\alpha}   & F_{\alpha f_{\rm evo} }\\
     F_{f_{\rm evo} \alpha} & F_{f_{\rm evo} f_{\rm evo}}\\
    \end{pmatrix}\;,
\end{equation}
where $F_{\alpha\alpha}$ is calculated in equation~\eqref{eq:Fisher_element}, while $F_{\alpha f_{\rm evo}}$ and $F_{f_{\rm evo} f_{\rm evo}}$ are 
\bea
F_{\alpha f_{\rm evo}} &=& \sum_i \frac{V_i}{4 \pi^2}\int \dd k k^2   \frac{\partial P_1 \left( k ,z_i\right)}{\partial \alpha} \left(\frac{\partial P_1 \left( k ,z_i\right)}{\partial f_{\rm evo}}\right)^* \sigma^{-2}_{P_1} \left( k,z_i \right) \;,\\
F_{f_{\rm evo} f_{\rm evo}} &=& \sum_i \frac{V_i}{4 \pi^2}\int \dd k k^2  \left| \frac{\partial P_1 \left( k ,z_i\right)}{\partial f_{\rm evo}} \right|^2 \sigma^{-2}_{P_1} \left( k,z_i \right) \;.
\eea
The derivative of the dipole $P_1$ with respect to $f_{\rm evo}$ can be computed analytically since the only dependence comes from the factor $\mathcal{R}$. By using that $ \frac{\partial \mathcal{R} }{\partial f_{\rm evo}}=-1$ we then obtain
\be
\frac{\partial P_1 \left( k ,z_i\right)}{\partial f_{\rm evo}}=-i f \frac{\mathcal{H}}{2k}P(k,z)\;,
\ee
where the factor $1/2$ arises from the assumption $b_A\left( z \right)- b_B\left( z \right)=0.5$.
Consequently, one can obtain the expected error for the parameter $\alpha$ marginalizing over $f_{\rm evo}$ for the case when both dark matter halos and galaxies are comoving with $f_{{\rm evo},A}=f_{{\rm evo},B}=0$, obtaining
\be \label{eq:Delta_alpha_marg_fevo}
\Delta\alpha=2\sqrt{{F}^{-1}_{\alpha\alpha}}=0.242 \;.
\ee
As we can see in Fig.~\ref{Fig:alpha_fevo}, $\alpha$ and $f_{\rm evo}$ are strongly degenerated in any single redshift bin. However, due to the redshift evolution, the direction of degeneracy rotates as we change from one bin to another, thus leading to the much smaller error-bar~\eqref{eq:Delta_alpha_marg_fevo}, once we account for the range of redshifts given in Table~\ref{tab:densities_bias}.

On one side our Fisher matrix approach may slightly under-estimate the degeneracy since we assume all the redshift bins to be independent and the fiducial model is set to $f_{{\rm evo},A}=f_{{\rm evo},B}=0$. On the other side, we have not assumed any prior on $f_{\rm evo}$ that we will realistically have for any upcoming surveys~\cite{Camera:2014bwa,Maartens:2021dqy}.

\section{Galaxy motion through geodesic equation}
\label{app:geodesic}

In this appendix we want to consider the case where the velocity of galaxies is fully determined by the geometrical perturbations, by solving the geodesic equation for a massive body
\be
\frac{d^2 x^\mu}{ d\tau^2}+ \Gamma^\mu_{\alpha \beta} \frac{dx^\alpha}{d\tau} \frac{dx^\beta}{d\tau} =0 \,,
\ee
where $\tau$ is the proper time defined by
\be \label{eq:normalization_cond}
g_{\mu \nu} \frac{dx^\mu}{d\tau} \frac{dx^\nu}{d\tau} =-1 \, .
\ee
Our system of coordinates are defined as $\left( x^\mu \right) = \left( t, r, \theta, \phi \right)$, where $t$ is the conformal time.
Therefore we need to determine
\bea
\ndv &=& \bn \cdot \bv = -v^r = - \frac{dr}{dt}= -\frac{dr}{d\tau} \frac{d\tau}{dt} \, ,\\
\dot \ndv &=& - \partial_t v^r \, .
\eea
At the background level we have
\bea
\frac{d^2 \bar t}{d\tau^2 } + \HH \left[ \left( \frac{d\bar t}{d\tau} \right)^2 +\left( \frac{d\bar r}{d\tau} \right)^2 +  \bar r^2 \left( \frac{d \bar \theta }{d\tau} \right)^2+\bar r^2 \sin^2 \bar \theta \left( \frac{d \bar \varphi }{d\tau} \right)^2 \right] &=&0 \, , \\
\frac{d^2 \bar r}{d\tau^2 } + 2 \HH \frac{d\bar t }{d\tau} \frac{d \bar r}{d\tau} -\bar r \left( \frac{d\bar \theta}{d \tau}\right)^2 - \bar r \sin^2 \bar \theta \left( \frac{d\bar \varphi }{d\tau} \right)^2 &=&0 \, , \\
\label{eq:geod_theta_0}
\frac{d^2 \bar \theta}{d\tau^2 } + 2 \HH \frac{d\bar \theta}{d \tau} \frac{d\bar t}{d\tau} + \frac{2}{\bar r} \frac{d\bar \theta}{d\tau} \frac{d \bar r }{d\tau} - \cos \bar \theta  \sin \bar \theta \left( \frac{d\bar \varphi }{d\tau}\right)^2 &=&0 \, , \\
\label{eq:geod_phi_0}
\frac{d^2 \bar \varphi}{d\tau^2 } + 2 \HH \frac{d\bar \varphi}{d \tau} \frac{d\bar t}{d\tau} + \frac{2}{\bar r} \frac{d\bar \varphi}{d\tau} \frac{d \bar r }{d\tau} + \cot \bar \theta    \frac{d\bar \varphi }{d\tau} \frac{d \bar \theta }{d \tau} &=&0 \, .
\eea
As expected from isotropy and homogeneity, $\frac{d\bar \theta}{d \tau} \equiv \frac{d \bar \varphi}{d\tau} \equiv 0$ are solutions of the differential equations~(\ref{eq:geod_theta_0}-\ref{eq:geod_phi_0}). So at the background we need only to solve
\bea
\label{eq:geod_t_0_v2}
\frac{d^2 \bar t}{d\tau^2 } + \HH \left[ \left( \frac{d\bar t}{d\tau} \right)^2 +\left( \frac{d\bar r}{d\tau} \right)^2 \right] &=&0 \, , \\
\frac{d^2 \bar r}{d\tau^2 } + 2 \HH \frac{d\bar t }{d\tau} \frac{d \bar r}{d\tau}   &=&0 \, .
\eea
The latter equation leads to
\be
\frac{d \bar t}{d\tau} \frac{d \bar u^r}{d \bar t} + 2 \HH \frac{d \bar t}{d\tau} \frac{d \bar r}{d\tau} =0 
\Rightarrow 
\frac{d \bar u^r}{d \bar t} + 2 \HH \bar u^r =0 \Rightarrow \bar u^r = \frac{C_1}{a^2} \, .
\ee
By plugging in this into the first differential equation we obtain
\be
\bar u^t \frac{d \bar u^t}{d \bar t} + \HH \left[ \left( \bar u^t \right)^2 + \frac{C_1^2}{a^4} \right] =0 
\Rightarrow 
\bar u^t = \pm \frac{\sqrt{a^2 C_2+C_1^2}}{a^2} \, .
\ee
From the normalization condition~\eqref{eq:normalization_cond} we get $C_2 = 1$
such that
\be
\left( \bar u^\mu \right) = \left( \pm \frac{\sqrt{a^2 + C_1^2}}{a^2}, \frac{C_1}{a^2}, 0,0 \right)
\ee
and the peculiar velocity
\be
v^r = \frac{dr}{dt} = \frac{u^r}{u^0} = \frac{C_1}{\sqrt{C_1^2 +a^2}} +\mathcal{O} \left( \epsilon \right) \, .
\ee
Due to background homogeneity we should set $C_1=0$.
So we have 
\be
\left( \bar u^\mu \right) = \left( a^{-1},0,0,0 \right) \, .
\ee
Going on to first order in perturbation theory we have from the normalization condition~\eqref{eq:normalization_cond} 
\be
u^0 = \frac{1- \Phi}{a}.
\ee
By solving the geodesic equation for $\delta u^r$ we obtain
\be \label{eq:vr}
v^r = \frac{\delta u^r}{\bar u^0} = \frac{a_{\rm IN}^2}{a} \delta u^r_{\rm IN} - a^{-1} \int_{t_{\rm IN}}^t  a \partial_r \Phi \left( t', r \left( t' \right) \right) dt'
\ee
where the integral runs over the galaxy path.
Then, the acceleration of the radial velocity is determined by
\bea
\dot \ndv &=& - \partial_t v^r =  \frac{a^2_{\rm IN}}{a} \HH \delta u^r_{\rm IN} - \frac{\HH}{a} \int_{t_{\rm IN}}^t  a \partial_r \Phi \left( t', r \left( t' \right) \right) dt'+ a^{-1} \partial_t  \int_{t_{\rm IN}}^t  a \partial_r \Phi \left( t', r \left( t' \right) \right) dt'
\nonumber \\
&=&
 \frac{a_{\rm IN}^2}{a} \HH \delta u^r_{\rm IN} - \frac{\HH}{a} \int_{t_{\rm IN}}^t  a \partial_r \Phi \left( t', r_* \right) dt'+  \partial_r \Phi \left( t,r_* \right) 
 \nonumber \\
 &=& \HH \left( \frac{a_{\rm IN}^2}{a}  \delta u^r_{\rm IN} - a^{-1} \int_{t_{\rm IN}}^t  a \partial_r \Phi \left( t', r_* \right) dt' \right) + \partial_r \Phi
 \nonumber \\
 &=& \HH v_r + \partial_r \Phi = -\HH \ndv +\partial_r \Phi
\eea
where we have set $r\left( t \right) = r_* $ (at background). 
This fully agrees with the Euler equation in $\Lambda$CDM, see eq.~\eqref{eq:Eulerradial}. But it is important to remark that in this derivation we never assumed any particular model of dark matter or dark energy. Therefore in the limit of galaxy being treated as test particle in a gravitational field, their motion is fully determined by the geometry through the equivalence principle.

\bibliography{dipole_test}
\bibliographystyle{JHEP}
\end{document}